\documentclass[twocolumn,showpacs,preprintnumbers,amsmath,amssymb,prb]{revtex4}
\usepackage{amsmath}
\usepackage{graphicx}% Include figure files
\usepackage{color}% Include colors
\usepackage{dcolumn}% Align table columns on decimal point
\usepackage{bm}% bold math

\begin{document}
\title{Coherent Control of Floquet-Mode Dressed Plasmon Polaritons}

\author{Regine Frank}%
%\email{regine.frank@kit.edu}
\affiliation{Institut f\"ur Theoretische Festk\"orperphysik, Karlsruhe
  Institute of Technology (KIT), Wolfgang - Gaede - Strasse 1, 76131
  Karlsruhe, Germany
}           

\date{\today}

\begin{abstract}
We study the coherent properties of plasmon polaritons optically excited on periodic nano-structures. The gold grains are coupled to a
single mode photonic waveguide which exhibits a dramatically reduced
transmission originating from the derived quantum interference. With a non-equilibrium description of Floquet-dressed polaritons we
demonstrate the switching of light transmission through the waveguide due to
sheer existence of intraband transitions in gold from right above the
Fermi level driven by the external laser light. 
\end{abstract}


\maketitle

\section{INTRODUCTION}

One of the ultimate challenges of optical technologies is the realization 
of efficient optical switching and computing
devices \cite{Ctistis, Koos, Kirschner}. Therefore,
quantum-optical functional elements have received increasing interest over the past years.
Among the promising candidates, quantum-well structures exhibiting Kerr
nonlinearities, meta-materials 
or plasmonic systems \cite{Gjonaj, Lav, Fleischer} have shown 
the potential to be used for ultrafast optical switches
\cite{Giessen_nat, Wegener, Woggon, Garcia}, and expeditiously progressing ultrashort laser sciences open
an avenue to their exploration \cite{Stockman, Hommelhoff_1}.
Besides its immense technological importance, 
the theoretical description of externally driven non-equilibrium quantum systems 
is a challenge itself.

\section{SETUP}

We study a photonic silicon-on-insulator hollow core waveguide (SOI) in contact with gold
nano-grains exposed to an external field. The
external time-periodic field modifies the electronic band structure within the
nano-grains by generating photo-induced Floquet-bands which can be attributed to
the Franz - Keldysh effect \cite{Dittrich}. All metals exhibit intraband
transitions occurring within the conduction band featuring a more or less small
absorption rate, whereas the governing processes are the interband
transitions. In gold (Au) a significantly different behaviour has been found
and attributed to the specific electronic structure of closed packed Au,
namely to the high polarizability of the $5d^{10}$
cores \cite{Whetten}. The collective resonance exhibits a large red shift to approximately
$2.4 eV$ which leads to the fact that the corresponding intensity is governed
by the interband but its sheer existence results from the
occurrence of intraband transitions which can be recognised as step like
structures in the spectrum \cite{Daniel}. These transitions, which result from a non fermi alike
distribution of states, have been observed in pump probe experiments with Au nano-spheres of diameters below $30
nm$. Their occurrence has been interpreted as the border from bulk like characteristics to the quantum regime of nano-particles. 
We show that the pump laser modifies the single plasmon band which results in
the development of Floquet-bands. Both frequency and amplitude
of the external field allow easy and fast-switching control over the position
in energy of these bands and sensitively control the generation
of a Fano-resonance with the photonic SOI mode. We show that this
switching may significantly alter the transmission properties within the
SOI by the formation of polaritons, electron photon bound states, at the
surface of the nano-grains. Consequentially, the potential of this quantum-optical functional element for all
optical switching based on polaritonics is proven. 

\begin{figure}[t!]
\begin{center}
  \scalebox{0.52}[0.52]{\includegraphics[clip]{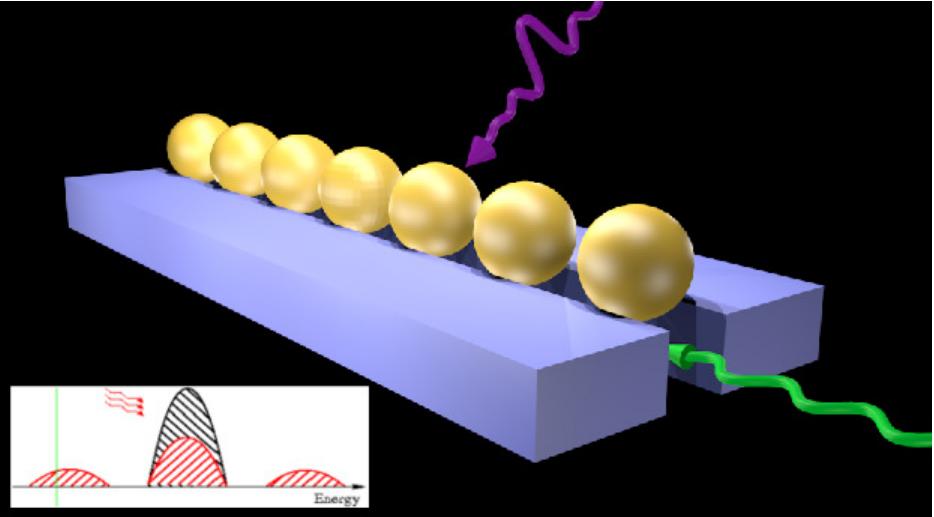}}\end{center}
\caption{(Color online) Gold nano-grains in contact with a hollow core
  SOI. Waveguide photons (green) and electrons in the metal form a coupled light-matter
state, a polariton controlled by an 
external laser (pink). Inset: Sketched development of Floquet side-bands. 
}
\label{Fig_00}
\end{figure}

\section{MODEL AND METHODS}

Our setup (Fig. \ref{Fig_00}) is described by a Fr\"ohlich Hamiltonian for
fermion-boson interaction
which has to be solved by applying the Keldysh formalism with respect to the
non-equilibrium character of the considered processes on the femto-second time
scale. We consider the SOI in 
contact with nano-grains, the nano-grain in the external field and finally discuss the non-equilibrium solution of the complete system in 
terms of electron Keldysh - Green's function and SOI transmission. As our
starting point we choose a single-band electronic tight binding model with
nearest-neighbour hopping \cite{Pauli}, 
characterised by the hopping amplitude $t$ , with the dispersion for a cubic
lattice $\epsilon_k= 2
t \sum_i \cos (k_i a)$, $a$ is the lattice constant and $k_i$ are the
components of the wave-vector. We assume a SOI supporting a single mode
$\hbar\omega_0 \!\!= \!\!2.34eV$. The SOI itself shall
be coated and therefore not being exposed to laser radiation. The electrons
may couple with strength $g$ weakly  to
SOI photons with
 frequency $\omega_0$. Hence the full Hamiltonian reads
\begin{eqnarray*}
H\!&=&\!\! \sum_{k, \sigma} \! \epsilon_k  c^{\dagger}_{k,\sigma}c^{{\color{white}\dagger}}_{k,\sigma} 
+ \!\! \hbar\omega_o  a^{\dagger}a^{{\color{white}\dagger}}
\!\\&& + g\! \sum_{k, \sigma}
c^{\dagger}_{k,\sigma}c^{{\color{white}\dagger}}_{k,\sigma}  (a^{\dagger}  \! +
a) - t\!\! \sum_{\langle ij \rangle, \sigma}\!\!
c^{\dagger}_{i,\sigma}c^{{\color{white}\dagger}}_{j,\sigma}  
\\&& + i\vec{d}\cdot\vec{E}_0 \cos(\Omega_L \tau)\sum_{<ij>} 
 \left(
           c^{\dagger}_{i,\sigma}c^{{\color{white}\dagger}}_{j,\sigma} 
 	  -
           c^{\dagger}_{j,\sigma}c^{{\color{white}\dagger}}_{i,\sigma} 	  
 \right).
\end{eqnarray*}

\begin{figure}[t!]
\begin{center} \scalebox{0.09}[0.09]{\includegraphics[clip]{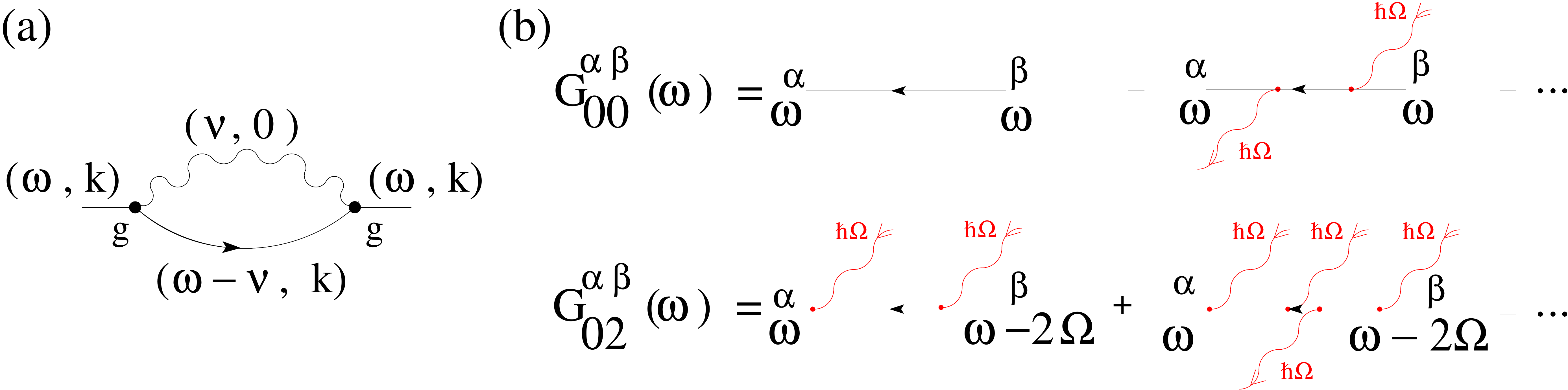}}\end{center}
\caption{
(Color online)(a): Contribution to the electronic self-energy $\Sigma (\omega, k)$ in second order perturbation
theory. The photon propagator does not carry a momentum as explained in the
text. 
(b): Floquet Green's function in terms of absorption/emission of 
external energy quanta $\hbar\Omega_L$.
$G^{\alpha\,\beta}_{00} (\omega)$, $\alpha , \beta$ being Keldysh indices, e.g. 
represents  the sum of all balanced processes. 
$G^{\alpha\,\beta}_{02} (\omega)$ describes a net-absorption of two photons.
}
\label{Fig_Selfenergy}
\end{figure}

\begin{figure}[t!]
\begin{center} \scalebox{0.35}[0.35]{\includegraphics[clip]{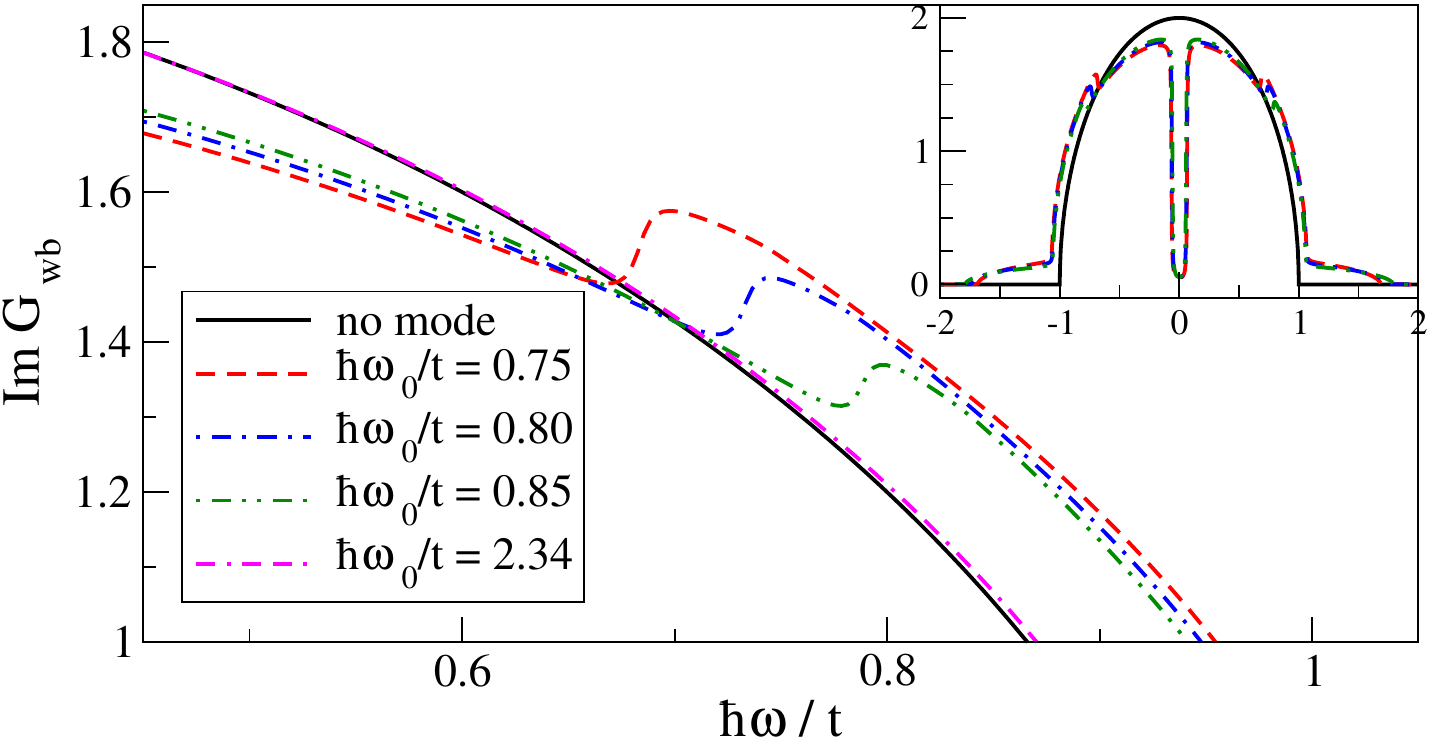}}\end{center}
\caption{
(Color online) The electron's spectral
function for various frequencies  of the SOI mode for coupling strength
$(\!g/t)^2\!\! =\! 0.09$ (about $\!30\%\!$ of the coupling $A_0/t$) 
at zero temperature for a spectral width $\tau=0.005$ of the
waveguide mode. Shape and position changing of the resonance is
clearly visible. If the photonic mode is energetically far off the electronic
band, as for $\hbar\omega_0\!=\!2.34eV$, the electronic band structure remains practically 
unchanged. The inset shows the overall behaviour of  spectral function.
}
\label{Fig_1}
\end{figure}

For a clear description we treat first the setup without the external
laser field, than the interaction of laser and band electrons, and finally the
full compound of external laser radiation interacting with band electrons and the
coupling of the latter by a Fano resonance with the waveguide mode. The Hamiltonian without the external laser field reads, 
\begin{eqnarray}
\label{Hamilton_we}
H\!=\!\! \sum_{k, \sigma} \! \epsilon_k  c^{\dagger}_{k,\sigma}c^{{\color{white}\dagger}}_{k,\sigma} 
+ \hbar\omega_o  a^{\dagger}a^{{\color{white}\dagger}}
\!+g\! \sum_{k, \sigma}
c^{\dagger}_{k,\sigma}c^{{\color{white}\dagger}}_{k,\sigma}  (a^{\dagger}  \! +
a).
\end{eqnarray}
Here, we assume the spatial extension of the Au nano-grains to be small ($< 30 nm$)
compared to the wavelength of the photonic mode inside the
SOI. Therefore, the momentum of the photons is much less than the
electron's momentum and we can set $q_{\rm photon} \simeq 0$ whenever we consider
the electronic subsystem. Thus, $a^{\dagger}$ ($a$) does not carry an index. 
In Eq. (\ref{Hamilton_we}), $ \epsilon_k$ is the electronic band energy,
$c^{\dagger}_{k,\sigma}$ ($c^{{\color{white}\dagger}}_{k,\sigma}$) creates
(annihilates) an electron with momentum $k$ and spin $\sigma$. 
$\hbar\omega_0  a^{\dagger}a^{{\color{white}\dagger}}$ is the photon energy eigen-state,
where $a^{\dagger}$ ($a$) creates (annihilates) a photon inside the SOI
with energy $\hbar\omega_0 $. The last (coupling) term on the r.h.s. is the standard term 
resulting from  
coupling the electronic and the photonic subspaces.
Due to the weak interaction between the SOI photons and the electrons
in the nano-grains we treat this interaction perturbatively.
In second order a self-energy contribution as shown in Fig. \ref{Fig_Selfenergy} is obtained from  Eq. (\ref{Hamilton_we}).
The coupling of the electronic system with a continuous energy spectrum to the
photons with a discrete one leads to a Fano resonance, which is observed in the electronic density of states,
as demonstrated in Fig. \ref{Fig_1}. Here we show the electron's spectral
function for various frequencies of the SOI mode for coupling strength
$(\!g/t)^2\!\!=\!\! 0.09$ at zero temperature for a spectral width
$\tau=0.005$ of the
waveguide mode (measured in units of the hopping $t$) at half filling, 
yielding  a suppression of the spectral function around the Fermi-level (half width $\tau$) 
where electrons are transferred to the high (low) energy tails of the spectral function. 
We note, that if the energy of the SOI mode $\hbar\omega_0$ is
distinctly different from the energy $\hbar\omega$ of the band-electrons, the
electronic density of states remains unchanged.

The subsystem of a nano-grain exposed to a
semiclassical electromagnetic laser field is described by the Hamiltonian (Lb = laser + band-electrons) 

\begin{eqnarray}
\label{Hamilton}
H_{Lb}\!=\! - t\!\! \sum_{\langle ij \rangle, \sigma}\!\!    c^{\dagger}_{i,\sigma}c^{{\color{white}\dagger}}_{j,\sigma} 
 \!+\! H_{C}(\tau)
\end{eqnarray}
where $\langle ij \rangle$ implies summation over nearest neighbores.  
$H_{C}(\tau)$ represents the  coupling to the external, time-dependent 
laser field, described by 
the electric field $\vec{E}=\vec{E}_0\cos(\Omega_L \tau)$, via the electronic dipole
operator $\hat{d}(\vec{x})$ with strength $\vec{d}$. It is given by
 \begin{eqnarray}
 \label{def_H_A}
 H_{C} (\tau) = i\vec{d}\cdot\vec{E}_0 \cos(\Omega_L \tau)\sum_{<ij>} 
 \left(
           c^{\dagger}_{i,\sigma}c^{{\color{white}\dagger}}_{j,\sigma} 
 	  -
           c^{\dagger}_{j,\sigma}c^{{\color{white}\dagger}}_{i,\sigma} 	  
 \right). 
\end{eqnarray}

The Hamiltonian specifically describes the excitation of a plasmon-polariton,
which corresponds to spatially delocalized intraband electronic motion caused by an external
electromagnetic wave. The accelerating energy is
immediately transferred into the motion of electrons by means of single-band
nearest neighbour hopping without interaction between the electrons.
Due to the time dependence of the external field,  Green's functions truly depend
on two separate time arguments. Therefore, we use a double Fourier
transform from time- to frequency space introducing relative and
center-of-mass frequency

\begin{eqnarray}
\label{Floquet-Fourier}
G_{mn}^{\alpha\beta} (\omega)
\!\!\! &=&\!\!\!
\left\lmoustache \!\!{\rm d}{\tau_1^\alpha}\!{\rm d}{\tau_2^\beta}\right.
e^{-i\Omega_L(m{\tau_1^\alpha}-n{\tau_2^\beta})}
e^{i\omega({\tau_1^\alpha}-{\tau_2^\beta})}
G (\tau_1^\alpha,\tau_2^\beta)\nonumber\\
\!\!\!&\equiv&\!\!\!
G^{\alpha\beta} (\omega-m\Omega_L, \omega - n\Omega_L),
\end{eqnarray}

where $(m,n)$ label the Floquet modes and $(\alpha, \beta)$ specify on which branch
of the Keldysh contour ($\pm$) the respective time argument resides. 
In general, a Floquet state is the analog to a Bloch state: The first
one results from a time-periodic potential whereas the latter is the
result of a space-periodic potential and both induce a band-structure.
A physical interpretation of 
such a Keldysh - Floquet Green's function 
is given  in Fig.\ref{Fig_Selfenergy}.
The special case of non-interacting electrons allows an analytical solution for 
$G_{mn}(k,\omega) $
by solving the equation of motion. 
Including photo-induced hopping,   
 the exact retarded Green's function for this sub-system reads
\begin{eqnarray}
G_{mn}^{R}(k,\omega) 
=
\sum_{\rho}
\frac
{
J_{\rho-m}\left(A_0\tilde{\epsilon}_k \right)
J_{\rho-n}\left(A_0\tilde{\epsilon}_k \right)
}
{
\omega -\rho\Omega_L - \epsilon_k + i 0^+
}
\end{eqnarray}
where $\tilde{\epsilon}_k$ represents the dispersion relation induced by the
external field Eq. (\ref{def_H_A}) and is different from $\epsilon$ Eq. (\ref{Hamilton_we}).
The $J_n$ are the cylindrical Bessel functions of integer order, 
 $A_0 = \vec{d}\cdot\vec{E}_0 $ and
$\Omega_L$ is the laser frequency.
The physical Green's function is given according to
\begin{eqnarray}
\label{EqGsum}
G^{R}_{\rm Lb}(k,\omega)  
=
\sum_{m,n}
G_{mn}^{R}(k,\omega).
\end{eqnarray}
\vspace*{0.1cm}

\section{RESULTS AND DISCUSSION}

\begin{figure}[t!]
\scalebox{0.36}[0.34]{\includegraphics[clip]{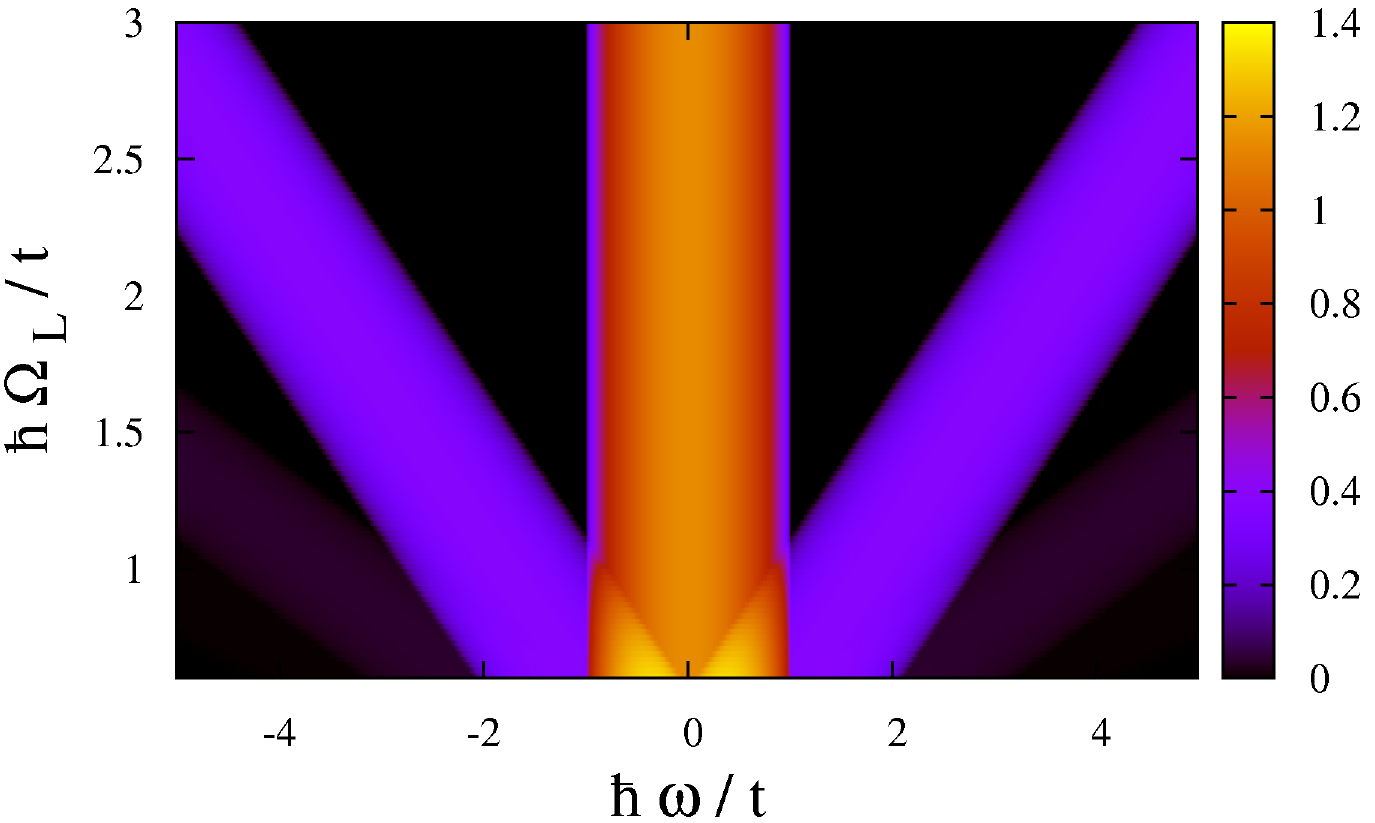}}
\caption{
(Color online) The imaginary part of the local Green's function,
Eq. (\ref {EqGsum}), is displayed as a function of quasiparticle energy
$\hbar\omega/t$ and external frequency $\hbar\Omega_L/t$ at zero temperature for external
amplitude $A_0/t=2.5$. 
The original semicircular DOS evolves side-bands as the
laser frequency 
increases. Sidebands of first (bright fuchsia) and second order (faded violet) can be identified.
}
\label{Fig_1b}
\end{figure}

\begin{figure}[t!]
\scalebox{0.36}[0.36]{\includegraphics[clip]{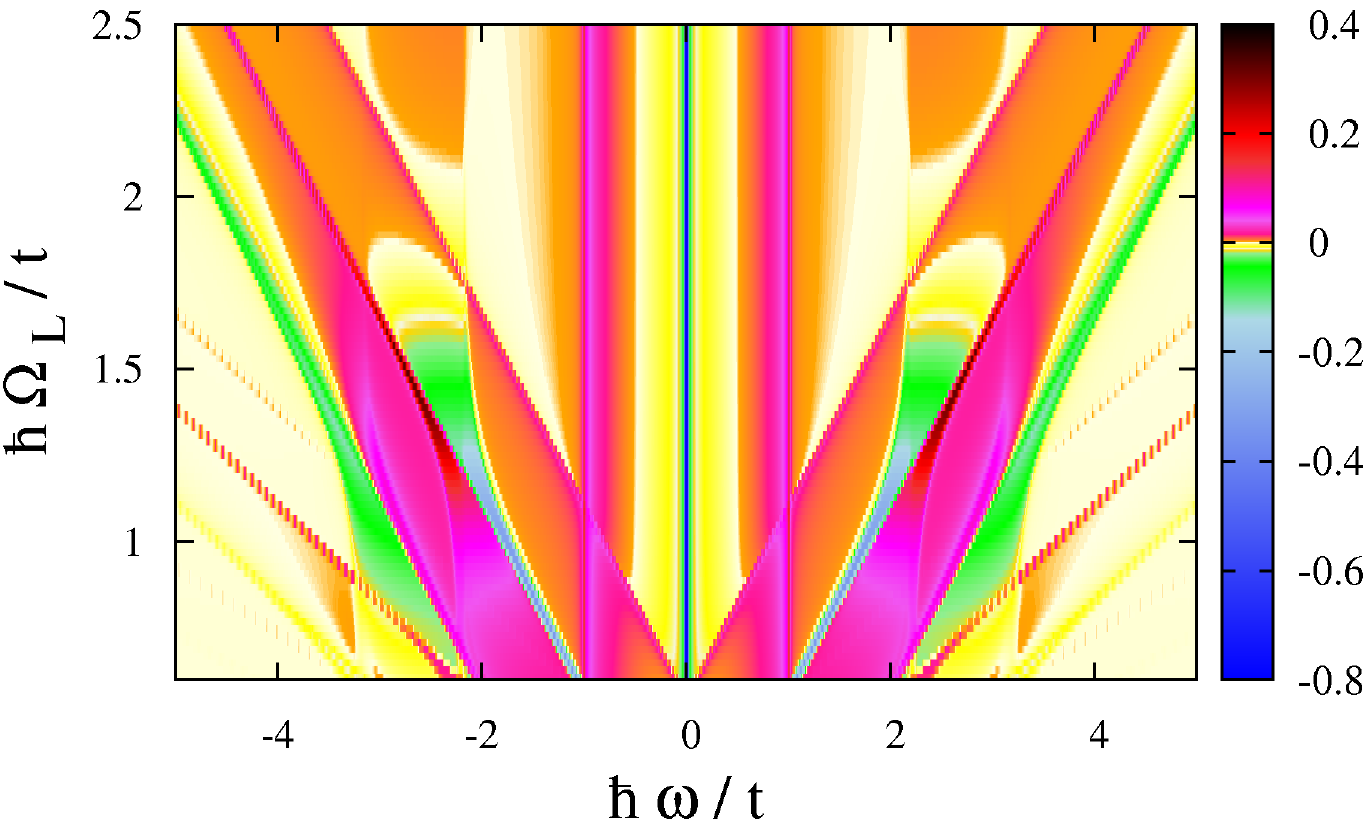}}
\caption{
(Color online) Laser induced change of the electronic density of
states $\delta G (\omega, \Omega_L) $.
A Fano resonance around quasiparticle energies $\hbar\omega_0 \!\!=
\!\!2.34eV$ is found, as soon as the external laser field redistributes the  electronic spectral weight such, that
the SOI mode finds electrons with about the same
energy to efficiently interact with, namely to get absorbed. 
}
\label{Fig_4}
\end{figure}

We present a numerical evaluation of Eq. (\ref{EqGsum}) in
Fig. \ref{Fig_1b}, where  ${\rm Im}\,G^{R}_{\rm Lb}(k,\omega)$, is displayed as a function of quasiparticle energy
$\hbar\omega$ and external frequency $\Omega_L$ at zero temperature for
$A_0/t=2.5$. 
As a typical value for the hopping we chose $t=1eV$. The SOI is operated
at the frequency $\hbar\omega_0=2.34 eV$ which corresponds to a frequency
doubled Nd-YAG laser ($\hbar\omega_0=1.17 eV$). The external
laser shall be characterised by $10 fs$
pulses, and shall be $E_0=6.299\times 10^{9} V/m$ ($I\sim 5.266 \times 10^{12} W/cm^2$)
in the surface region of the Au grains, including Mie type
field enhancement effects due to
the small particle sizes. For the
nano-grains we choose a damage threshold of $0.5 J/cm^2$ and $|d|=6.528\times 10^{-29}Asm$ ( with the lattice constant for Au $a_{Au}=
4.08 \times 10^{-10}m$) 
resulting in $A_0 \equiv d \cdot E_0=2.5 eV $ and the peak of the
electronic response
characteristics is developed after $0.3 ps$ after pulse injection, whereas the
damping sets in right afterwards and
so does the bleaching which is reduced with decreasing particle size \cite{Orrit}. Temperature effects are significantly lowered for
nano-size particles and the observed transmission change can be
totally attributed to the electronic response \cite{cool}. For later use, we
also assume a density of $50\%$ nano-grains per $10$ $\mu m$ of the SOI.
The original semicircular density of states develops
photonic side-bands, the bandstructure, as the external laser frequency
$\Omega_L$ increases.
Because of  the
point-inversion symmetry of the underlying lattice, the first side-band
represents the two-photon processes, the less pronounced second
side-band the four-photon processes.
Their occupation is described by the non-equilibrium distribution function as
calculated from the Keldysh component of the Green's function.
\vspace*{-0.1cm}
In a last step, we combine the relaxation processes due to the interaction
between the band electrons and the SOI as described by
Eq. (\ref{Hamilton_we}), with the impinging external laser as introduced
in Eq. (\ref{Hamilton}).
The resulting Green's function consequentially describes the SOI
with Au nano-grains that themselves are now exposed to the external laser
radiation. We treat the weak coupling between the electrons and the SOI
photons by second order perturbation theory and the interaction between the
electrons and external laser in terms of the Floquet theory, as demonstrated above.
Since we are interested in possible switching effects, we choose as the
initial situation the case where the photonic mode, $\hbar\omega_0 = 2.34 eV$,  
is far off the electronic bandedge. Electrons arrive
band energies ranging from $-1\le\hbar\omega/t\le+1$.
A Fano resonance is only observable in the presence of external radiation
of appropriate frequency, i.e. only if the induced side-bands meet 
the energy of the SOI mode.
This resonance will also affect the transmission properties in the SOI,
since now SOI photons can be absorbed in the formation of a
mixed state of light and matter with the laser-induced charge excitations in the
Au nano-grains. Thus a waveguide polariton is created yielding to a
significant reduction of the SOI's transmission.
In Fig. \ref{Fig_4}, we display the laser induced change of the density of states $\delta G$ as a function
of quasi-particle energy $\hbar \omega$ and external laser frequency $\Omega_L$.The quantity $\delta G $ measures

\begin{figure}[b!]
\begin{center} \scalebox{0.32}[0.32]{\includegraphics[clip]{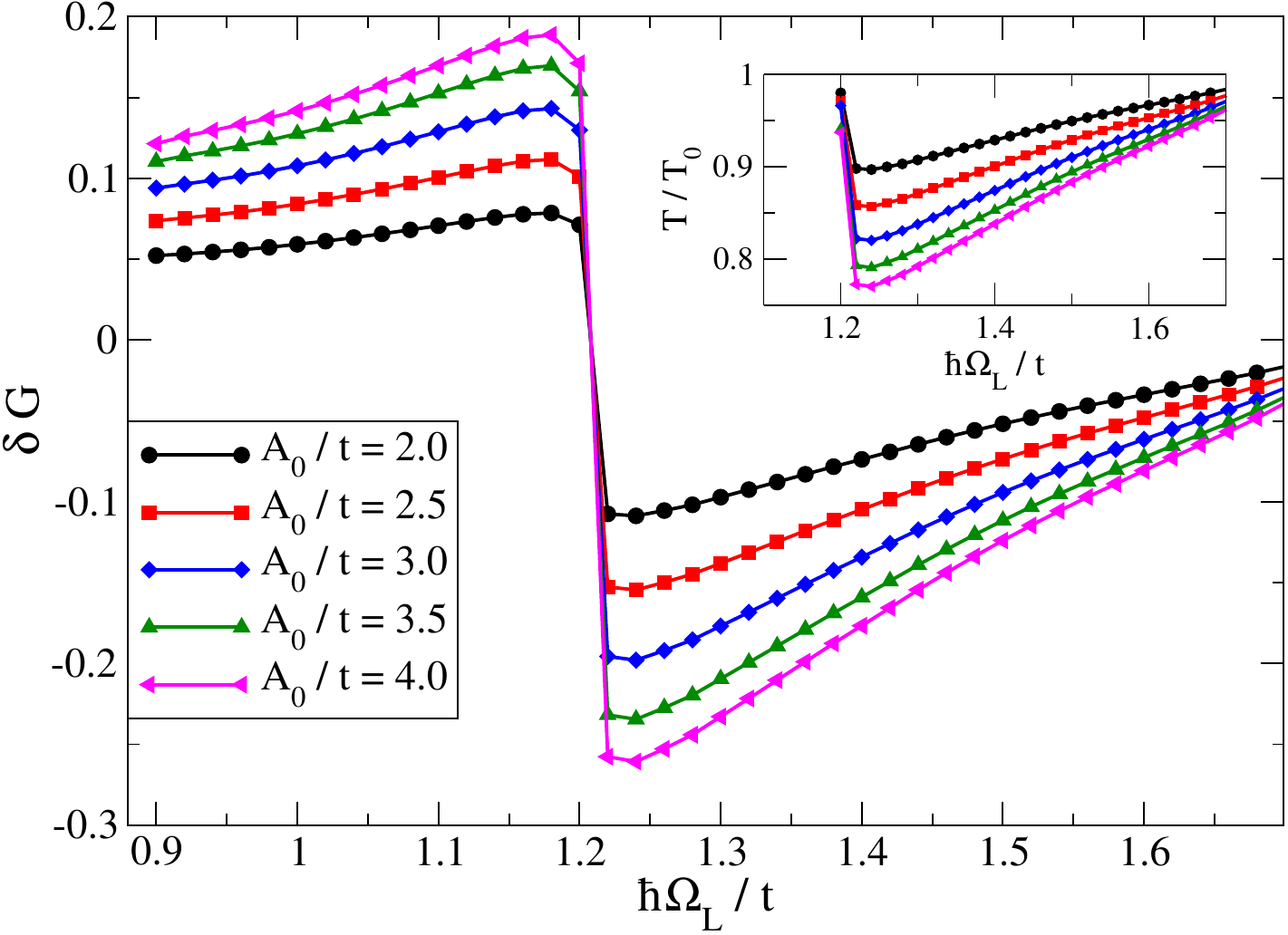}}\end{center}
\caption{
(Color online) Laser induced change of the electronic
DOS, $\delta G $, at fixed quasiparticle energy which meets the discrete value
of the SOI mode
$\hbar\omega\!\!=\!\!2.34 eV\!\!=\!\!\hbar\omega_0$. Inset: Relative photon transmission 
in a SOI of unit length $l\!\!=\!\!l_o$ as a function of 
laser frequency $\Omega_L$.
}
\label{Fig_2}
\end{figure}

\begin{eqnarray}
\label{delta_G}
\delta G 
&=&
\left[ 
{\rm Im\,} G_{\rm } (\omega, \Omega_L)
-
{\rm Im\,} G_{\rm Lb} (\omega, \Omega_L)
\right]\\
&&-
\left[ 
{\rm Im\,} G_{\rm wb} (\omega)
-
{\rm Im\,} G_{\rm b} (\omega)
\right]\nonumber
\end{eqnarray}
the effect of the impinging laser field on the electronic density of
states, and vanishes as the external laser and the coupling to the
SOI is turned off. In Eq. (\ref{delta_G}),  $G_{\rm }$ represents
the Green's function including all processes, $G_{\rm Lb}$ the interaction
between the laser field and the band electrons as given in
Eq. (\ref{EqGsum}), $G_{\rm wb}$ describes the SOI in presence of the
band electrons and is solution to Eq. (\ref{Hamilton_we}), and finally 
$ G_{\rm b}$ is the Green's function of just the noninteracting band electrons.
In Fig. \ref{Fig_4} the laser induced change of the electronic density of
states $\delta G (\omega, \Omega_L) $ experiences a Fano resonance when the
external laser redistributes electronic spectral weight leading to the
absorption of the SOI mode at $\hbar \omega_0 = 2.34 eV$. That behaviour is derived when the first photonic
side-band meets the energy of the SOI mode yielding a sign change in $\delta G$ at this energy.
In Fig. \ref{Fig_2}, the laser induced change of the electronic
density  of states $\delta G $ is shown at fixed quasiparticle energy $\hbar
\omega=\hbar \omega_0$,
where $\omega_0$ is the frequency of the SOI mode. 
Asymptotically, i.e., for large $\Omega_L$, $\delta G$
vanishes, as already indicated in Fig. \ref{Fig_1b}, because in this limit
there is no electronic spectral weight at the energy of the SOI mode.
In the opposite limit, $\Omega_L \rightarrow 0$, the influence of the laser
field is non-zero, because here higher-order laser induced side-bands exist,
yielding spectral weight at the resonance position already for smaller laser
frequencies. That result can also be concluded from the second-order side-band in
Fig. \ref{Fig_1b}. In a SOI of length $l$, the ratio between the initial and
the transmitted intensity is given by $T\sim \exp(-\alpha\, l/l_o)$. Here $\alpha / l_o$,  is the absorption coefficient divided by the unit length
$l_o$, where $\alpha$ includes an average over one period of the external
periodic driving field with frequency $\Omega_L$. 
We recognise that $\omega\delta G $ can be understood as the
leading contribution to the relative absorption coefficient as discussed in
detail in
ref. \cite{Jauho_1}. 
The relative transmission of photons $T/T_0$
within the SOI of unit length $l=l_o$ is shown in Fig. \ref{Fig_2} as a function of the external
laser frequency $\Omega_L$. 
System parameters are given in
caption of Fig. \ref{Fig_4}.
Depending on the frequency of the driving field, an intensity drop of up to
$25 \%$ is observed (Fig. \ref{Fig_2}), and by varying the length of the SOI the
transmission inside the SOI can in fact be turned on and off. 

\section{CONCLUSION} 

We have presented a quantum field theoretical model
for a SOI in contact with gold nano-grains
which themselves are exposed to external laser irradiation. 
The strong and coherent external laser is  described in terms of 
the Floquet theory, assuming classical behaviour of this oscillatory-in-time
field, whereas the interaction with the SOI mode reflects a quantum
interference. Such a description has never been
proposed before to the best of our knowledge and the obtained results demonstrate the high potential  of
SOI polaritons for all-optical switching. Both, frequency and amplitude of the external laser control transmission through the SOI, 
and each of these features ensure ultrafast switching processes.\\

\section{ACKNOWLEDGEMENTS} 

The author thanks A. Lubatsch, K.-P. Bohnen, K. Busch, F. Hasselbach, W. Nisch,
A. v. Raaij, G. Sch\"on and M. Wegener for various
fruitful discussions. P. Hommelhoff and his group are greatfully
acknowledged for highly efficient correspondence.
The author is mentor of Karlsruhe School of Optics \& Photonics (KSOP) and
thanks for support and funding.

\end{document}